\documentclass[A4]{article}
\usepackage[total={6in,8.5in}]{geometry}

\usepackage[authoryear]{natbib}
\usepackage{hyperref}
\usepackage{amsmath}
\usepackage{amssymb}
\usepackage{lmodern}
\usepackage{graphicx}
\usepackage{authblk}

\usepackage[capitalize]{cleveref}
\creflabelformat{equation}{#2#1#3}

\usepackage{color}
\usepackage[dvipsnames]{xcolor}

\definecolor{changecolor}{rgb}{.9,0,0} 
\definecolor{lukacolor}{rgb}{0.1,0.5,0.15} 
\definecolor{vincenzocolor}{rgb}{0,0.6,0.9}

\usepackage[colorinlistoftodos,%
]{todonotes} 

\title{Bayesian Detection of Mesoscale Structures\\in Pathway Data on Graphs}

\author{
Luka V Petrovi\'c$^{\star}$ and
Vincenzo Perri$^{\star}$
\\
{
Data Analytics Group,\\
Department of Informatics,\\
University of Zurich,\\
Zurich, Switzerland\\
$\star$ equal contributions
}
}

\begin{document}
\maketitle

\begin{abstract}
    Mesoscale structures are an integral part of the abstraction and analysis of complex systems.
    They reveal a node's function in the network, and facilitate our understanding of the network dynamics.
    For example, they can represent communities in social or citation networks, roles in corporate interactions, or core-periphery structures in transportation networks.
    We usually detect mesoscale structures under the assumption of independence of interactions.
    Still, in many cases, the interactions invalidate this assumption by occurring in a specific order.
    Such patterns emerge in pathway data; to capture them, we have to model the dependencies between interactions using higher-order network models.
    However, the detection of mesoscale structures in higher-order networks is still under-researched.
    In this work, we derive a Bayesian approach that simultaneously models the optimal partitioning of nodes in groups and the optimal higher-order network dynamics between the groups.
    In synthetic data we demonstrate that our method can recover both standard proximity-based communities and role-based groupings of nodes.
    In synthetic and real world data we show that it can compete with baseline techniques, while additionally providing interpretable abstractions of network dynamics.
\end{abstract}

\section{Introduction}
\label{hog:intro}

The identification of mesoscale structures in a network is a cornerstone of the analysis of complex systems across domains.
The basic premise is to identify the groups of nodes in the network that are similar, and analyse group interactions.
Since the number of groups will be much smaller than that of nodes,  the group interactions provide a coarse-grained description of the network's dynamics that is easier to understand and to analyze than the interactions between nodes in their full resolution.
Two main types of mesoscale structures are communities and roles.
Communities are groups of nodes in the network that are more likely to interact within the group than with the nodes from other communities.
In other words, detecting communities can be seen as detecting nodes that are close to each other in the topology of network edges.
Roles are groups of structurally similar nodes, i.e., nodes belonging to the same role have similar properties, interact with nodes of similar kinds, and need not to be close in the network topology.

Networks represent complex systems as independent dyadic interactions between nodes.
However, interactions between nodes can be more complex: they can depend on each other in the order in which they appear.
These dependencies in the temporal ordering of interactions can be captured by pathway data, which consists of sequences of nodes that some process has traversed.
If there are no dependencies between interactions, the next node on a sequence only depends on the current node, and not on any previous one.
If there are dependencies, the next node on a sequence will depend on the longer history of interactions in the path, and not just the current node.
To capture such dependencies, researchers have modeled pathway data with higher-order network models, a type of higher-order Markov models tailored to networks, and found that this can improve the analysis~\citep{lambiotte2019networks}.

\begin{figure*}[t]
    \centering
    \includegraphics[width = \textwidth]{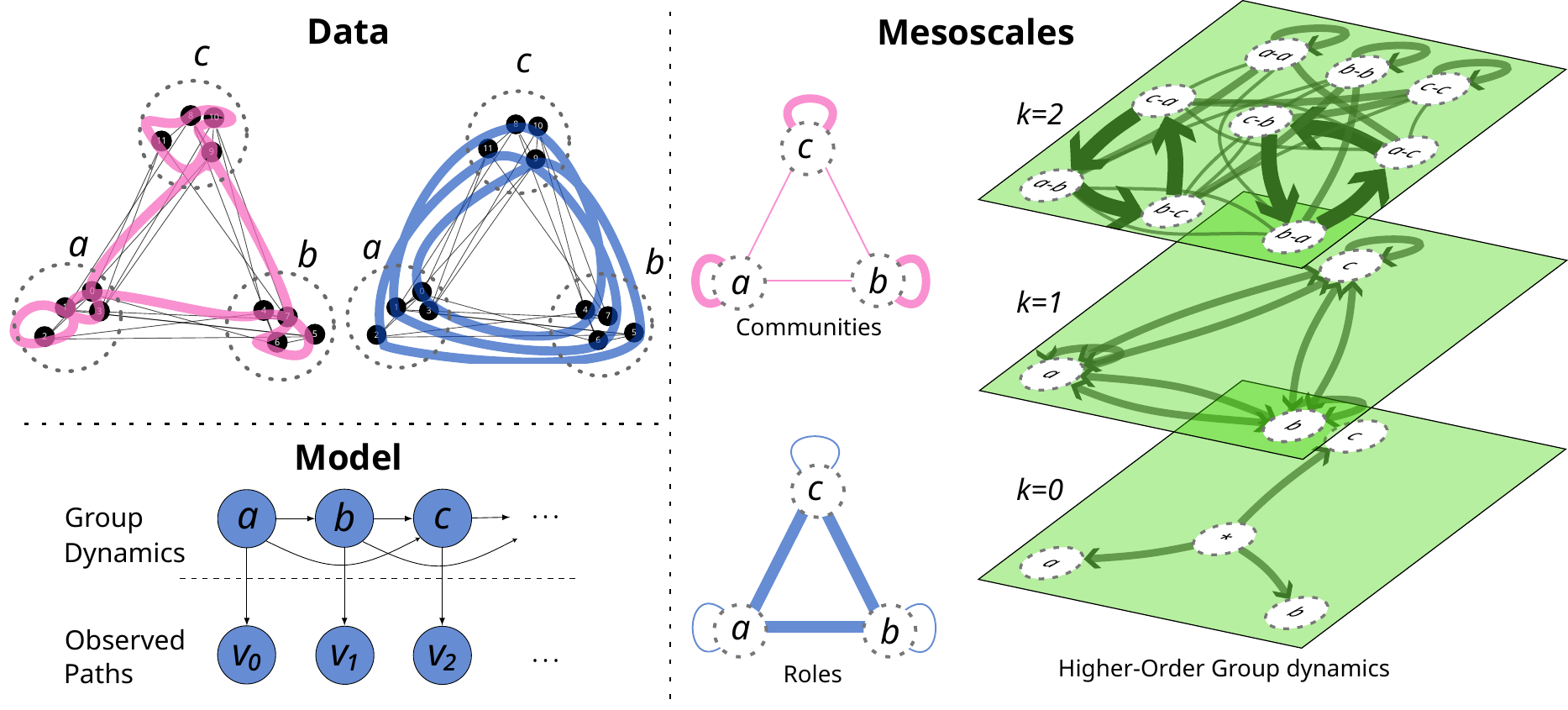}
    \caption{
    Top left: examples of empirical paths for two different pathway dynamics in a network of $12$ nodes and $3$ groups ($a$, $b$, and $c$).
    From the perspective of a standard network model, the two dynamics are a community pattern (pink) and a role pattern (blue). 
    A higher-order approach would additionally notice that the blue path also has memory---it flows circularly through groups (clockwise or counterclockwise).
    Bottom left: the generative model for paths that we use is based on the assumption that the path generating process walks through groups based on a higher order dynamics and chooses nodes based on the current group.
    Right: Two examples of first-order group dynamics and one example of second-order group dynamics.
    A group dynamics of a (first-order) community structure is shown in pink: each member of a community has a higher chance to connect with the members of the same community.
    A group dynamics of a (first-order) role structure is shown in blue: each role has a higher probability connecting to nodes of different roles.
    A second-order group dynamics is presented as three green layers. 
    Bottom layer shows the probability to start a path in different groups, the first order layer shows the dynamics of the first transition, and the second layer shows the dynamics of all remaining steps on a path. 
    One can clearly see two cycles, indicating the patterns $abca$ and $cbac$ of higher-order group dynamics that generate the blue path.
    }
    \label{fig:teaser}
\end{figure*}

Mesoscale structures in higher-order networks have been investigated mostly as a problem of community detection~\citep{rosvall2014memory,peixoto2017modelling}.
The communities are identified as groups of nodes where the process (e.g. a walker) tends to stay in~\citep{bovet2022flow}.
However, the dynamics of paths can be such that paths go across groups of nodes, e.g. at a company, the driving pattern may be the interactions between people with different roles.
Thus, the relevant mesoscale structures would resemble the roles, more than the communities.
To our knowledge, this broader view of mesoscale structures in higher-order networks has not been investigated.
Therefore in this work, we take a wider look at the dynamics of groups of nodes, and investigate how to detect nodes that behave similarly in sequences representing paths in graphs.

In this work:
\begin{enumerate}
    \item we derive a Bayesian approach to simultaneously detect the optimal mapping of nodes to groups and the optimal higher-order dynamics between groups;
    \item we demonstrate in synthetic data with a known ground truth that the approach detects both the communities and roles, and that it retrieves the correct dynamics between groups;
    \item we show the practical value of our method in real-world data, comparing it with the node embedding technique node2vec\citep{grover2016node2vec} and the community detection algorithm InfoMap\citep{rosvall2008maps}.
\end{enumerate}

We organize the article as follows.
In \cref{hog:problem formulation}, we describe our assumptions about the data and the goals we want to achieve by analyzing such data.
In \cref{hog:related work}, we discuss related work and describe the differences to our approach, which is explained in \cref{hog:methods}.
In \cref{hog:experiments}, we use synthetic data with known ground truth to explore the mesoscale structures that can be detected with our method: we first look at communities and roles, then at more general higher-order patterns.
In this section, we also compare our approach with alternatives in synthetic and real-world data, discuss the strengths and limitations of our work, and the possible societal impact.
Finally, we conclude in \cref{hog:conclusion} and summarize the opportunities for future work.

\section{Problem Formulation}
\label{hog:problem formulation}
\newcommand{\nodes}{{V}}
\newcommand{\edges}{{E}}
\newcommand{\graph}{{{G}}}
\newcommand{\groups}{\mathcal{G}}
\newcommand{\groupmap}{{\gamma}}
\newcommand{\probabilitysimplex}[1]{{\Delta^{#1}}}
\newcommand{\Dirichlet}{{\mathrm{Dir}}}
\newcommand{\Beta}{{\mathrm{B}}}
\newcommand{\p}{{\pi}}
\newcommand{\maxorder}{{K}}
\newcommand{\hmon}{{\mathcal{M}}}
\newcommand{\emissionmodel}{{\mathcal{E}}}
\newcommand{\hog}{{\mathcal{H}}}
\newcommand{\successors}{{S}}
\newcommand{\paths}{{P}}

This section explains the basic assumptions about the data and the goals we want to achieve with our technique.

We assume that we are given a set of nodes $v\in\nodes$, and a multiset of paths $\paths$ containing paths $(v_0, v_1, \ldots, v_l)$ of potentially variable length $l$.
In addition, we potentially assume that the paths are a result of some process running on a network $\graph=(\nodes, \edges)$, where $\edges \subset \nodes \times \nodes$.
The paths would thus have to obey the network topology, i.e. a path can contain a transition $v_i$ to $v_{i+1}$ only if $(v_i, v_{i+1}) \in \edges$.
This condition would reduce the degrees of freedom of our models, and thus improve its data efficiency, however, we note that this assumption does not reduce the generality of our approach, as one can always assume that the graph is fully connected, resulting in absence of constraints. 
Some examples of datasets satisfying these conditions are trajectories of cars in a city, documents being passed between company employees, passengers in a transportation network, and web users generating clickstream data.

Having such data, our first goal is to detect groups $\groups$ of nodes with similar dynamical behavior in pathway data.
This can be a community structure, where a path that starts in one group of nodes stays within the same group for some time, e.g. a web user that reads several articles about popular science before moving on to politics. 
It can also be a role structure, where a path has to go through some nodes with a certain function before continuing forward, e.g. a document in company that has to be approved by someone from the legal department.
We thus operationalize "node behavior" as patterns of (higher-order) transition probabilities in paths. 

\section{Related work}
\label{hog:related work}

Our problem can be seen as a specific type of hidden Markov model inference. 
The community investigating Hidden Markov models (HMM)~\citep{rabiner1986introduction} assumes the existence of hidden labels (states) that influence the observable sequences of complex systems.
The inference of HMMs~\citep{leroux1992maximum,stolcke1993hidden,bicego2003sequential,cappe2006inference,sarkar2018bayesian} is a topic of scientific investigation to date.
The problem we address is similar to the inference of HMMs, since the groups that we are interested in correspond to hidden states, and the nodes in paths to observed states in sequences.
However, our problem setting is, on the one hand, simpler than for general HMMs because the groups that we consider are disjoint sets (whereas a hidden state could in the general case trigger an emission of any observable state) and on the other hand, more complex than for general HMMs, because we assume that we are given large number of short sequences, and not a single long sequence.
Aside from that, our work has a common theme of topological constraints that was investigated in the continuous setting by \citet{roweis2000constrained}.
We also note that \citet{smyth1997clustering} used HMMs to cluster variable-length sequences, albeit the sequences are made of vectors in continuous spaces.

The second body of literature pertains to the partitioning of nodes of a static graph, commonly referred to as node clustering.
The problem has been widely investigated in the network science literature, but mostly for the detection of communities, i.e. groups of nodes with a higher density of connections within the group than outside the group~\citep{fortunato2016community}.
However, we can cluster nodes according to criteria that are different from the density of their connections.
This is particularly true for directed networks, where nodes can be clustered together based on the similarity of connectivity patterns that do not require them to share connections \cite{malliaros2013clustering,rossi2014role}.
This notion of clusters is akin to the detection of roles.
In \citep{jin2019smart}, the $k$-step neighborhood of email exchanges in a workplace is used to extract node features that are used for identifying work-roles.
Similarly, in \citep{henderson2012rolx} features that capture the node's role in the network are combined in a vector representation of nodes.
Building on these ideas, we explore node clustering in dynamic graphs, where temporal dependencies between edges offer more information about the community and role structures.

Both \citet{henderson2011rolx,jin2019smart} use vector representation of nodes for role detection, which highlights the connections of our work to the literature on embeddings, i.e., the learning of vector representations of network nodes. 
These vector representations place nodes in a Euclidean space such that nodes with a similar connectivity pattern in the network are closer in the embedding space.
Given the vector representation of nodes, node clusters can be obtained by applying clustering algorithms for Euclidean spaces on the representations of nodes.
\citet{perozzi2014deepwalk} obtain an embedding by sampling neighborhoods through random walks.
Similarly, \citet{grover2016node2vec} sample the network with a biased random walk and show that, depending on the type of bias, the resulting node embedding can represent either communities or roles.
Strengthening the connection of embedding methods and mesoscales, \citep{rossi2020proximity} proposes a taxonomy of network embedding approaches that categorizes methods according to whether their representations focus on representing community or role patterns.
Our work attemts to generalize these patterns, using a model that can record both communities and roles.
Additionally, in embeddings, it is often unclear what network patterns lead to similar or dissimilar node representations.
Instead, our model not only groups nodes but also has an interpretable representation of the their mesoscale dynamics.

The last body of literature relevant to our work deals with identifying mesoscale structures in higher-order networks.
The non-Markovian patterns that characterize higher-order networks are shown to lead to cluster structures that are different from the ones expected from the static analysis of networks~\citep{rosvall2014memory}.
In addition, \citet{scholtes2014causality} highlights that higher-order clusters not only alter, but can also be in contrast with the static clusters. 
\citet{edler2017mapping} propose a hierarchical extension of InfoMap \cite{rosvall2008maps} that accounts for higher-order patterns by applying the algorithm to node sequences.
\citet{bovet2022flow} cluster nodes in temporal networks based on whether walkers are likely to remain within their starting cluster.
\citet{benson2015tensor} extend matrix-based spectral approaches for networks partitions to tensors.
Through this extension, they define the proximity between higher-order network structures and obtain a partition that minimizes the number of cuts of a specified higher-order (role) pattern.
\citet{peixoto2017modelling} propose a Bayesian technique that, for a sequence of nodes, jointly clusters memories and nodes, and finds its Markov order.
The problem they consider is similar to the detection of dynamic clusters addressed in this work. 
However, there are two significant differences: first, we cluster only the nodes, and detect a higher-order dynamic between groups; second, our model is specially tailored for paths on graphs, meaning that it can deal with a large number of short paths and take network constraints as input.

\section{Higher-Order Group Detection}
\label{hog:methods}
In this section, we mathematically describe our Bayesian approach: we give the formulas for the likelihood, explain how we learn the posterior of model parameters from the data, and how we perform model selection.

The method is based on predicting the next node $v_i$ that will be visited by the path $(v_0, v_1, \ldots, v_{i-1})$ , i.e. we want to approximate the distribution
\begin{equation}
    \label{hog:eq:predictnextnode}
    p(v_i | v_0, \ldots, v_{i-1})
\end{equation}
We assume that each node $v$ belongs to one of disjoint groups $g\in\groups$.
Since our aim is to statistically learn the groups $\groups$ and their higher-order dynamics, we model the probability distribution in \cref{hog:eq:predictnextnode} using two modelling assumptions: (1) the nodes appear on the paths only based on their group label, and independent of other factors (2) the group that is visited next by the path is dependent exclusively on the last $\maxorder$ steps of the path.
Formally:
\begin{equation}
    \label{hog:eq:factorization}
    p(v_i | v_0, \ldots, v_{i-1}) = p(v_i|g_i) p(g_i|g_{i-\maxorder}, \ldots, g_{i-1})
\end{equation}
This factorization splits the problem into two parts: modelling the emission probabilities $p(v_i|g_i)$, and modelling the group dynamics $p(g_i|g_{i-\maxorder}, \ldots, g_{i-1})$.
These two parts together give us a complete generative model for paths, which is very similar to HMMs, barring that the groups $\groups$ are disjoint.
The factorization is also similar to the MapEquation~\cite{rosvall2008maps}, with the difference that our definition of groups does not expect the paths to remain within a group for some time before exiting. 
The emission model captures the activity of nodes within groups, while the group dynamics defines the higher-order dependencies between interactions. 
In the following, we explain each of those two parts and the Bayesian model selection of the whole model.

\subsection{The Emission Probabilities}
\label{hog:methods:em}
An emission model models the emission of a node $v$ given its group $g$ with a categorical distribution with parameters $p(v|g) = \p_{v|g}$.
Each parameter $\p_{v|g}$ is dependent on other parameters $\p_{w|g}$ from the same group $w\in g$, since $\sum_{v\in g} \p_{v|g} = 1$.
Vectors $\vec{\p}_g = (p_{v|g})_{v\in g}$ are independent, and belong to the probability simplex $\vec{\p}_g \in \probabilitysimplex{|g|-1}$.
Thus, the number of emissions $n_v$ of each node $v$ from a group $g$ is modeled by a multinomial distribution with parameters $\vec{\p}_{g}$, and the probability to see $n_v$ times each node $v\in V$  is:
\begin{equation}
    \label{hog:eq:emission:likelihood}
    p((n_v)_{v\in \nodes}| \vec{\p}_{g} ) = \prod_{g\in\groups} \prod_{v\in g} \p_{v|g}^{n_v}
\end{equation}

\subsection{Bayesian Inference of the Emission Probabilities}
\label{hog:methods:bem}

Instead of working with point estimates of $\vec{\p}_{g}$, we employ Bayesian learning and calculate the distribution of parameters $\vec{\p}_{g}$.
Therefore, for the emission model $\emissionmodel$, we a priori assume that the distribution of each vector $\vec{\p}_g$ is uniform:
\begin{equation}
    \label{hog:eq:emission:prior}
    p( (\vec{\p}_{g})_{g\in\groups}|\emissionmodel) = \prod_{g\in\groups} \Dirichlet(\vec{\p}_{g}|\vec{\alpha}_{g}^0)
\end{equation}
where $\vec{\alpha}_g = (\alpha_{v|g}^0)_{v\in g}$ and where $\alpha_{v|g}^0 = 1$ for all $v \in g$, corresponding to the uniform prior, and $\Dirichlet$ denotes Dirichelet distribution.

Having observed pathway data $\paths$ with $n_v$ observations of each node $v$, the posterior distribution is computed using the Bayes rule:
\begin{equation}
    p((\vec{\p}_{g})_{g\in\groups} | \paths, \emissionmodel) = \frac{p(\paths|(\vec{\p}_g)_{g\in\groups}, \emissionmodel ) p((\vec{\p}_{g})_{g\in \groups}| \emissionmodel )  }{p(\paths | \emissionmodel)}
\end{equation}

Importantly, the denominator (called ``marginal likelihood'') can be integrated analytically:
\begin{equation}
    p(\paths|\emissionmodel) = \prod_{g\in\groups} \frac{\Beta(\vec{\alpha}^0_g + (n_v)_{v\in g})}{\Beta(\vec{\alpha}^0_g)}
\end{equation}
where $\Beta$ is the well-known multivariate beta function.
Substituting \cref{hog:eq:emission:likelihood} and \cref{hog:eq:emission:prior}, we get the posterior analytically:
\begin{equation}
    \label{hog:eq:emission:posterior}
    p((\vec{\p}_{g})_{g\in\groups} | \paths) = \prod_{g\in\groups}\Dirichlet(\vec{\p}_{g}|\vec{\alpha}_{g})
\end{equation}
where $\vec{\alpha}_g = (\alpha_{v|g})_{v\in g}$ and $\alpha_{v|g} = \alpha_{v|g}^0 + n_v $ for all $v \in g$.

We note that there is only one model for a given group assignment, and thus there is no model selection for the emission model in isolation. 
We note, however, that the emission model in isolation would have higher marginal likelihood the more groups there are: in the extreme where each node is in its own group, any data would have the marginal likelihood of one.

\subsection{Group Dynamics}

To model the transitions between groups, we explore the Markov modelling techniques for pathway data.
Predicting the next step of a path based on last $\maxorder$ visited nodes has some peculiarities in comparison with standard sequences.
First, sequential data generally consists of a single long chain, while pathway data consists of large number of possibly short sequences.
This difference makes single-order Markov chains of order $\maxorder$ problematic because they do not model the first $\maxorder$ steps.
In contrast to a single long chain where the first few steps are a negligible part of the data, the first few steps of all paths might represent most of the pathway data.
Second, the paths on a graph do not permit an arbitrary node to follow any other node as some nodes might not be connected.
This means that the number of degrees of freedom of the system might not correspond to the number of degrees of freedom of a Markov chain, which can cause inefficiencies in the model selection. 
For those reasons, we resort to Multi-Order networks (MONs)~\citep{scholtes2017network} which are models specially tailored for pathway data, because they explicitly model the first few steps thus solving the first issue.
To deal with the second issue, we use the inference of MON models that can explicitly account for a graph constraint~\citep{petrovic2022learning}.
We give a brief description of MON models for completeness (a more detailed consideration and evaluation can be found in~\cite{scholtes2017network}), before explaining the inference in the \cref{hog:methods:bmon}.

Multi-order networks are made of $\maxorder + 1$ "layers" of higher-order network models of orders $k\in\{0,1,\ldots \maxorder\}$.
The zeroth-order layer describes the probability of starting a path in a group $g_0$, the first-order layer describes the probability of the first transition from group $g_0$ to a group $g_1$, and so on.
The first $\maxorder$ groups $g_i$ on a path $(g_0, g_1, g_2, \ldots, g_l)$ are modeled with the $i$-th layer, and all other steps are modeled with the $\maxorder$-th layer:
\begin{align}
    p(g_0, g_1, g_2, \ldots, g_l) = 
    &\prod\limits_{i=0}^{\maxorder-1} p(g_i|g_0, \ldots, g_{i-1})\times\nonumber\\ 
    &\prod\limits_{i=\maxorder}^{l} p(g_i|g_{i-\maxorder}, \ldots, g_{i-1}) 
\end{align}
For a history $\bar{g} = (g_{i-k}, g_{i-k+1}, \ldots, g_{i-1})$ of length $k\leq K$, and groups $g \in \successors(\bar{g}) $, where $\successors(\bar{g})$ denote possible successors of history $\bar{g}$, we denote the model parameters $p(g_i=g | \bar{g}) = \p_{g|\bar{g}}$.
The parameters  $(\p_{g|\bar{g}})_{g\in\successors(\bar{g})}$ that relate to the same history $\bar{g}$ are dependent on each other because $\sum_{g\in \successors(\bar{g})} \p_{g|\bar{g}} = 1$, but the vectors $\vec{\p}_{\bar{g}} = (\p_{g|\bar{g}})_{g\in\successors(\bar{g})}$ for each history $\bar{g}$ are independent of each other.

\subsection{Bayesian Inference of the Group Dynamics}
\label{hog:methods:bmon}

Bayesian inference of MONs and their model selection has been developed and evaluated in \citep{petrovic2022learning}.
It computes analytically, and is more data-efficient than the maximum likelihood approaches.
The inference is analogous to the Bayesian inference of the emission probabilities that we described in \cref{hog:methods:bem}, and thus we give a brief description of the Bayesian inference of MON models for completeness, but the reader can refer to \citep{petrovic2022learning} for the details.

Similar to the emission probabilities from a group, the transition probabilities from a history $\bar{g}$ of each layer $k$ to a group $g_i$ are also modeled with a multinomial distribution with parameters $\vec{\p}_{\bar{g}}$.
We model the distribution of the parameters $\vec{\p}_{\bar{g}}$ based on the counts $n_{g_i|\bar{g}}$ of the transitions.
For the multi-order network model $\hmon$, we assume the uniform prior distribution over each probability simplex:
\begin{equation}
    p((\vec{\p}_{\bar{g}})_{\bar{g}}|\hmon) = \prod_{\bar{g}} \Dirichlet(\vec{\p}_{\bar{g}}| \vec{\alpha}_{\bar{g}}^0 ) 
\end{equation}
where the a priori concentration parameters $\vec{\alpha}_{\bar{g}}^0 = (\alpha_{g|\bar{g}}^0)_{g\in\successors(\bar{g})} $ of the Dirichlet distribution are $\alpha_{g|\bar{g}}^0 = 1$ for all $g\in \successors(\bar{g})$.
The posterior can be calculated analytically in the same spirit as for the emission model.
Assuming that we observed paths $\paths$ containing $n_{g|\bar{g}}$ counts of transitions from history $\bar{g}$ to a group $g$, the posterior is:
\begin{equation}
    p(\vec{\p}_{\bar{g}} | \paths, \hmon )  = \Dirichlet(\vec{\p}_{\bar{g}} | \vec{\alpha}_{\bar{g}} ) 
\end{equation}
where $\vec{\alpha}_{\bar{g}} = (\alpha_{g|\bar{g}})_{g\in\successors(\bar{g})}$ are concentration parameters, and the update rule is $\alpha_{g|\bar{g}} = \alpha_{g|\bar{g}}^0 + n_{g|\bar{g}}$.

The marginal likelihood is analogous to the marginal likelihood in the case of the emission model:
\begin{equation}
    p(\paths|\hmon) = \prod_{\bar{g}} \frac{\Beta(\vec{\alpha}^0_{\bar{g}} + (n_{g_i|\bar{g}})_{g\in \successors(\bar{g})})}{\Beta(\vec{\alpha}^0_{\bar{g}})}
\end{equation}
The model selection is performed by comparing the marginal likelihoods of different group dynamics.
To compare the models with two different maximum orders $K$ and $K'$, where $K<K'$, we note that we can assume that the simpler model ($K$) is the null model, and perform the hypothesis test using Bayes factors~\citep{kass1995bayes}, with the ``very strong'' threshold $B_{10} > 150$, which showed the best performance~\citep{petrovic2022learning}.
Thus, assuming we have the correct groups and enough data, we can also recover the correct group dynamics.
Comparing the different group assignments, we see that in contrast with the emission model, the marginal likelihood of the group dynamics model becomes $1$ when all nodes are assigned to the same group.
Thus the group dynamics balances the total marginal likelihood with the emission model, and allows us to detect the optimal number of groups.

The formula for the likelihood calculation of a single network partition $\groups$ can be computed analytically for any maximum order $\maxorder$.
For a given pathway dataset and order $\maxorder$, we require a single pass through the data, where we count the occurrences of each transition in the group space up to order $\maxorder$, to compute the likelihoods of each order $k<\maxorder$ and select the optimal one.

\subsection{Higher-Order Group Model}
\label{hog:methods:hog}

The higher-order group (HOG) model describes the transitions between nodes as a product of transition between groups and emission probabilities from groups to nodes.
It is akin to a hidden Markov model, with the differences that each hidden state can produce a disjoint set of observed states, and that the hidden process is a multi-order network model.
For given groups $\groups$, we define a group map $\groupmap(v) = g \Leftrightarrow v \in g$.
The higher-order group model consists of the probabilities of an emission model $\vec{\p}_{g}$ and the probabilities of a Multi-Order Network model of groups $\vec{\p}_{\bar{g}}$.
The likelihood of a path $(v_0, v_1, v_2, \ldots, v_l)$ is:
\begin{align}
    &~p(v_0, v_1, \ldots, v_l|\vec{\p}_{g},\vec{\p}_{\bar{g}} ) 
    = \prod\limits_{i=0}^{l} \p_{v_i|\groupmap(v_i)} \times \nonumber\\
    &\prod\limits_{i=0}^{\maxorder-1} \p_{\groupmap(v_{i}) | \groupmap(v_0), \ldots, \groupmap(v_{i-1})}
    \times \prod\limits_{i=\maxorder}^{l} \p_{\groupmap(v_{i}) | \groupmap(v_{i-K}), \ldots, \groupmap(v_{i-1})}
\end{align}
where the first product is for the emission model, and the other two are for the group dynamics for the first $\maxorder$ transitions and for the remaining transitions between the groups.

\subsection{Bayesian Inference of Higher-Order Group Model }

This section describes the Bayesian inference of higher-order groups (B-HOG).
We use the factorization of the probabilities (\cref{hog:eq:factorization}) and learn the emission model and the group dynamics separately and solely based on the group map $\groupmap:\nodes\rightarrow\groups$.
The marginal likelihood of the higher-order group model is also the product of marginal likelihoods of the emission model and the group dynamics.
For a group map $\groupmap$ and a dataset of paths $\paths$, where we have $n_v$ observations of each node $v$, and where have $n_{v|\bar{v}}$ observations of transitions from each history $\bar{v} = (v_1, v_2, \ldots, v_k)$ to each node $v$ the update rule is simply $\alpha_{v|g} = \alpha_{v|g}^0 + n_v$ for the emission model and $\alpha_{g|\bar{g}} = \alpha_{g|\bar{g}}^0 + n_{g|\bar{g}}$ for the group dynamics, where $\bar{g} = (g_1, \ldots, g_k)$ and  
\begin{equation}
    \label{eq:renameNodes2Groups}
    n_{g|\bar{g}} = \sum\limits_{\groupmap(v) = g, \groupmap(v_i)=g_i} n_{v|\bar{v}}. 
\end{equation}

We have to perform model selection in terms of finding the optimal group map $\groupmap$ and the optimal order $\maxorder$.
We discussed in \cref{hog:methods:bmon} how to perform a Bayes factor test between two different orders with the same group map.
To compare models with different group map $c_1$ and $c_2$, which are not nested, we simply choose the model with the highest marginal likelihood (i.e. $B_{10}>1$). 
Since the marginal likelihood is integrated, and not maximized, it does not suffer from overfitting \citep{mackay2003information}.

The investigation of strategies to explore the space of partitions was outside of the scope of this work since we are interested in the patterns that can be described, and since other authors addressed similar questions~\cite{peixoto2020merge}.
Instead, we use a simple Metropolis-Hastings procedure, where we draw the next candidate grouping by randomly changing one node's group, to explore the space of possible partitions.
We optimized the computational complexity of the search so that we have to pass through the full pathway data only once, which we explain in detail in \cref{hog:appendix:partitionSpaceExploration}.

\section{Experiments}
\label{hog:experiments}

We evaluate the application of B-HOG to both synthetic and empirical datasets.
In the following, we aim to answer three questions.
First, can B-HOG detect standard communities and roles from pathway data on networks? 
Second, can B-HOG detect mesoscale structures in pathways with higher-order patterns? 
Third, how does B-HOG perform in comparison to existing baselines for learning mesoscale structures on synthetic and empirical graphs?
We also demonstrate B-HOG's ability to infer interpretable group dynamics, which is an advantage in the analysis of complex data.

\paragraph{Baselines} 
We run a comparative analysis using  as baselines the embedding method node2vec \citep{grover2016node2vec}, and the information theoretic clustering algorithm InfoMap \citep{rosvall2008maps}.
Both baselines take weighted networks as input, which we construct by counting all transitions $(v_i,v_{i+1})$ from pathway data.
In our experiments, we trained node2vec with $80$ walks of length $40$, window of $10$, and for hyperparameters $ (p,q) \in [1,4] \times [1,4]$.
We apply the well-known clustering algorithm K-means \citep{lloyd1982least} on the learned vector representations setting the input number of clusters $k$ equal to the number of clusters in the ground truth.

\paragraph{Evaluation}
We evaluate detected groups by comparing to the ground truth.
For the evaluation, we use the adjusted mutual information (AMI)~\citep{vinh2009information}.
When AMI is zero, the detected groups do not correspond to the ground truth groups -- they are as good as random groups; when AMI is one, the detected groups correspond perfectly to the ground truth groups.

\paragraph{Synthetic experiments}
In our synthetic experiments, we aim to perform an exhaustive search over all group maps, and establish whether the global minimum of the marginal likelihood corresponds to the ground truth pattern. 
We create a network model with $n=9$ nodes and three groups $ g = \{a,b,c\}$.
In the first two examples, we analyse if B-HOG can recover the communities and roles, and we manually set the ground truth models. 
In the third experiment, we
analyse whether B-HOG can retrieve arbitrary pattern of mesoscale structures, and how it compares with the baselines, thus 
we sample ground truth models for paths with arbitrary higher-order patterns.
Because the number of nodes is small, we use a dimensionality $d =3$ for node2vec.

\paragraph{Synthetic Communities} First, we show that B-HOG detects standard community structures. 
Input paths are generated as random walks that have $70\%$ probability to move to a node in the same community and $15\%$ to move to a node from each of the remaining two groups.
We produce $500$ paths of length $10$.
Node2vec partially recovers the pattern ($\text{AMI}=0.34$) when we use a return parameter $p=1$ and in-out $q=4$.
InfoMap perfectly recovers the communities and obtains $\text{AMI}=1$. 
For B-HOG, we test all assignments between $1$ and $4$ groups, and up to $2$-nd order group dynamics. 
B-HOG identifies both the correct communities and their number ($\text{AMI}=1$) and the $1$-st order dynamics.
This group dynamics can be seen in \cref{fig:teaser} (right, pink).
It is characterized by dominant self-loops, because a path (shown in pink on bottom left) repeatedly visits nodes from the same group, before moving on to another one.

\paragraph{Synthetic Roles} 
Second, we show that B-HOG can detect roles in networks.
Input paths go between groups: from $a$ (or $b$, $c$) the path has a $45\%$ probability to go to $b$ or $c$ (respectively, $a$ or $c$, $b$ or $c$), and a $10\%$ probability to move to a node with the same role.
Node2vec partially recovers the pattern with $p=4$ and $q = 1$ obtaining and $\text{AMI}=0.33$, 
InfoMap obtains AMI of zero.
For B-HOG, we test all assignments between $1$ and $4$ groups, and up to $2$-nd order group dynamics.  
B-HOG correctly identifies both the role assignment ($\text{AMI}=1$) and the $1$-st order.
This group dynamics can be seen in \cref{fig:teaser} (right, blue).
It is dominated by connections between different groups, because a path (similar to the blue one on the left) visits in succession nodes from different groups.

\paragraph{Consistent detection of arbitrary synthetic dynamics} 
Third, we show that B-HOG can detect arbitrary mesoscale structures on graphs with higher-order dynamics.
We run $100$ independent experiments.
We randomly generate a ground truth group dynamics as fully connected directed network with uniformly random transition probabilities for the first order (\emph{synth-1}) and as a uniformly random MON model without constraints of orders $2$ (\emph{synth-2}) and $3$  (\emph{synth-3}).
This generates mesoscale structures that are a mix of communities and roles. 
For \emph{synth-2} and \emph{synth-3}, they also have higher-order dynamics.
We sample a $m = 10^5$ paths of length $10$, emit the nodes using a uniformly random emission model, and use the resulting paths as data.
We detect the groups with B-HOG, InfoMap and node2vec.
Results are presented in \cref{hog:table:synthetic_results}.
The baselines had variable performance, indicated by large standard deviation $\sigma$ of the AMI. 
B-HOG always identified the ground truth grouping and order as optimal, meaning that they are the global maximum of the marginal likelihood.
An example of a higher-order group dynamics can be seen in \cref{fig:teaser} (three green layers on the right).
Layer $k$ represent $k$-th layer of a MON model. 
The pattern is dominated by two cycles $abca$ and $cbac$ in the second order, because the path (shown in blue on the left) makes such cycles.
Higher-order group dynamics allow us to distinguish such patterns from static roles (shown in blue).

\begin{table}[h]
    \centering
    \begin{tabular}{ll|lll}
    dataset/number of experiments & Method & $k_\hog$ &  AMI & $\sigma$ \\
    \hline \hline 
    \emph{synth-1} & B-HOG & 1 &    1.0&0.0 \\
    {$100$ experiments}  & InfoMap & - &  0.11&0.31 \\
      & node2vec & - &  0.06&0.27 \\
    \hline 
    \emph{synth-2} & B-HOG & 2 &    1.0&0.0 \\
    {$100$ experiments}  & InfoMap & - &  0.04&0.22 \\
      & node2vec & - &  0.02&0.19 \\
    \hline 
    \emph{synth-3} & B-HOG & 3 &    1.0&0.0 \\
    {$100$ experiments}  & InfoMap & - &  0.0 &0.04 \\
      & node2vec & - &  0.15&0.28 \\
\end{tabular}      
    \caption{Synthetic experiments}
    \label{hog:table:synthetic_results}
\end{table}

\paragraph{Empirical datasets and Preprocessing} 
We use five real-world temporal networks collected by the SocioPatterns collaboration.
They contain time-stamped proximity interactions recorded at a resolution of $20$ seconds, and include information on the groups to which these nodes belong.  
Datasets \emph{school-11} and \emph{school-12} \cite{fournet2014contact} contain the time-stamped proximities between high-school students (126 students in 2011, 180 in 2012).
Students belong to classes, which we use as ground truth grouping.
Dataset \emph{hospital} \cite{vanhems2013estimating} contains interactions between patients and healthcare workers in a hospital ward. 
We use their roles (patient, nurse, administrative, doctor) as ground truth groups.
Datasets \emph{work-13} and \emph{work-15} \cite{genois2015data} capture interactions between employees recorded in an office building (92 workers in 2013, 217 in 2015).
Employees belong to departments, which we use as ground truth labels.  
Using the time-stamped interactions, we extract the pathways which are the input for our model. 
For each node $v$, we record the time-ordered sequence $p_v = (w_0, w_1, \ldots, w_l)$ of nodes $w \in V \setminus \{v\}$ that interacted with $v$.
Since some interactions last longer than 20 seconds, they are recorded in multiple consecutive time-stamps. 
We represent them as a single node in the sequence, as they represent a single interaction.

\paragraph{Group Dynamics in Empirical Datasets} 
In all empirical datasets, we provided the ground truth number of groups to all methods instead of only to node2vec (which uses it in $k$-means).
We tested B-HOG up to the $5$-th order dynamics.
For node2vec, we present here only the results for dimensionality $d=16$ and $p=1,q=1$; the results for the other hyperparameter choices performed similarly and can be found in \cref{hog:appendix:synthRes,hog:appendix:empRes}.
For B-HOG we ran $60 000$ iterations of the Metropolis-Hastings procedure.
The number of runs for B-HOG experiments is reported in \cref{hog:table:empirical_results}.
The results for the other two methods are obtained over more than $250$ runs.
For node2vec, we report the average and standard deviation of the runs. 
For InfoMap and B-HOG, we report the score obtained by the mapping with the highest log-likelihood (or shortest description length). 
The results are in \cref{hog:table:empirical_results}. 
We note that we detect higher-order dynamics between groups in all datasets.
Compared to InfoMap, our method scores worse in $3$ datasets, ties in $1$, and scores better in $2$, indicating that the community structures correspond well to the recorded node labels.
In comparison to node2vec our method scores worse in $2$ empirical datasets, ties in $1$, and scores better in the $3$ remaining ones.

\begin{table}[]
    \centering
    \begin{tabular}{ll|lll}
    dataset/runs & Method & $k_\hog$ &  AMI \\
    \hline\hline
    \emph{school-11} & B-HOG & 4 & 0.81  \\
    {$67$ runs} 
    & InfoMap & - & 0.79  \\
     & node2vec & - & 0.80 $\pm$ 0.02\\
     \hline 
    \emph{school-12} & B-HOG & 3 & 0.83  \\
    {$63$ runs} 
     & InfoMap & - & 0.76 \\
     & node2vec & - & 0.955 $\pm$  0.002 \\
     \hline 
    \emph{hospital} & B-HOG & 4 & 0.25  \\
    {$146$ runs} & InfoMap & - & 0.30  \\
     & node2vec & - & 0.10  $\pm$ 0.06\\
     \hline 
    \emph{primary} & B-HOG & 3 & 0.92  \\
    {$70$ runs}  & InfoMap & - & 0.93  \\
     & node2vec & - & 0.86 $\pm$ 0.02\\
     \hline 
    \emph{work-13} & B-HOG & 2 & 0.76  \\
    {$60$ runs} & InfoMap & - & 0.76  \\
     & node2vec & - & 0.76 $\pm$ 0.02 \\
     \hline 
    \emph{work-15} & B-HOG & 2 & 0.67  \\
    {$100$ runs} & InfoMap & - & 0.72  \\
     & node2vec & - & 0.70 $\pm$ 0.03
    \end{tabular}
    \caption{Empirical results}
    \label{hog:table:empirical_results}
\end{table}

\paragraph{Discussion} 
The results first and foremost indicate that B-HOG can recover the mesoscale patterns of communities and of roles.
Furthermore, we have seen that B-HOG can detect patterns that are more general than communities or role, and that it can recover the number of groups, groups themselves and their higher-order dynamics.
Our synthetic experiments also indicated that these generalized 
patterns can hardly be recovered with baseline techniques.
In empirical data, our method detected optimal orders $2$-$4$ in all data.
Since Bayesian learning has built-in Occam's razor these correlations are not spurious, and indicate the existence of higher-order patterns in empirical data.
To support these findings we perform additional experiments.
First, we note in \cref{hog:appendix:fixedorder} that we find lower description lengths with higher-order models than with first-order models in all data.
Second, we find in \cref{hog:appendix:optimalordergt} that the optimal order for ground truth labels, too, is $2$-$4$ for all datasets.
In terms of unsupervised recovery of the labels recorded in the empirical data, our method does not outperform the baselines.
This is not surprising, given that it detects more general patterns and the no free lunch theorem~\cite{peel2017ground,mccarthy2019exact}.
However, this does not imply that our method is not able to recover the mesoscale structures important for the dynamics. 
On the contrary, we argue that our method finds mesoscale structures that are even more important for the dynamics than the empirical labels.
To support this, in \cref{hog:appendix:optfromgt}, we perform an additional experiment where we use the ground truth labels as initial mapping and then try to find mappings with higher log-likelihood. 
As we always find maps with higher log-likelihood, the ground truth mappings do not compress the paths optimally, and thus, they do not model the dynamics of the system optimally.
Using our model, not only can we find groups that better model the paths, but we can also represent the dynamics of groups as a multi-order network, which can be further analyzed and interpretted using generalized network analytic tools~\citep{lambiotte2019networks,perri2021visualisation,gote2021predicting}.

\paragraph{Limitations} 

Our consistency analysis in synthetic data is empirical, and thus limited by design. 
In the future, we would try to establish a theoretical guarantee (similar to \citet{jiang2021consistent}).
The second limitation is the simple Metropolis-Hastings procedure, which limits the application to larger datasets. 
This limitation could potentially be addressed as it was for SBM~\citep{peixoto2020merge}.
The third possible limitation regards the choice between two models with different cluster assignments.
Since they are not nested, we simply selected the model with higher marginal likelihood.
However, the difference between the marginal likelihoods does not have to be significant, meaning that there could be several different models that fit the data reasonably well.
This could be solved by selecting all the models with marginal likelihood close to the best model, but it is mostly relevant for the cases where the data-size is small and thus was outside of the scope of the current work.

\paragraph{Societal Impact} 
We identified three opportunities for societal impact of our work.
First, we hope that our work would help researchers across the domains to analyse their pathway data in a simple and interpretable way. 
Our method produces abstractions of complex dynamic data on complex systems and provides another tool in the toolkit of applied researchers.
Second, our work contributes to a growing literature on higher-order network models~\citep{lambiotte2019networks}. 
It provides a solution to an unsolved problem of higher-order community detection and generalizes this concept to higher-order groups.
Third, we hope that the existence of techniques for higher-order analysis encourages researchers to collecting more data containing higher-order patterns, which would allow new insights into dynamics of various complex systems.
Since our method abstracts dynamic network data and helps us understand complex processes, the method can have positive or negative impact only insofar insights can be used to create positive or negative impact on society.
Having said that, we do not see any specific way in which our method can be misused to make a negative impact on society.

\section{Conclusion}
\label{hog:conclusion}

Analysing dynamic complex systems at a higher level of abstraction is a necessary step to understand them.
We can abstract complex networks by identifying mesoscale structures, however, we lack techniques for identifying mesoscale structure when complex networks are dynamic, and when network edges have temporal dependencies.
To this end, we propose a model that describes both the community partitions and the group dynamics.
Although identifying the optimal groups requires a search in the space of network partitions, for each partition the marginal likelihood of our model can be computed analytically.
We have shown that our method is able to automatically detect a more general class of mesoscale structures, which encompasses the standard communities and roles, and that it can additionally detect higher-order edge dependencies.
The group dynamics provides an abstraction of the dynamic network that is interpretable, and which can be analysed further to understand the system.

The main opportunity for future work is the investigation of the strategies to search the partition space, which is the major obstacle to applying our model to datasets with a large number of nodes.
The second opportunity lies in the investigation of the ensembles of node groupings.
Although the marginal likelihood of the model was simply maximized over the partition space, one could keep track of several best-performing groupings, or even sample them from the distribution. 
Our model would offer not only an ensemble of node groupings, but also of the dynamics of those groups.
If one could match the different dynamics, then one could also detect when nodes perform several roles in a complex system.
The third opportunity is to explore different priors of the group dynamics and investigate their effect on the analysis.
For example, if we construct the prior such that self-loops have higher concentration parameters than other transitions in the group dynamics, then, would the method be biased to detect communities in the system?
Similarly, if we amplify the prior concentration parameters of some (higher-order) transitions, then, would the method cluster nodes according to mesoscale structures corresponding to those transitions?
If true, our method would allow a researcher to choose a pattern and to ``view'' the data through the chosen pattern, while also being able to compare the significance of each perspective by computing the likelihood of the detected model.
The understanding of these research questions would build a versatile tool to analyze pathway data of dynamic complex systems, that could improve our understanding of mesoscale structures and their dynamics across domains.

\subsubsection*{Acknowledgements}
Authors thank Anatol Wegner and Lisi Qarkaxhija for the valuable comments on the manuscript and acknowledge support by the Swiss National Science Foundation, grant 176938.

\bibliography{bibliography}

\appendix

\section{Exploration of the Partition Space}
\label{hog:appendix:partitionSpaceExploration}

\newcommand{\ngroups}{{n_G}}
We use a simple Metropolis-Hastings procedure to explore the space of possible partitions.
For a given set of nodes $\nodes$, and a chosen maximum number of groups $\ngroups$, we assign each node to a random group.
For a given dataset $\paths$ and current partition $\groups$, we create a next candidate partition $\groups'$ by selecting (uniformly at random) a single node $v$, and assigning to it a different group label $\{ 0,\ldots, \ngroups-1 \}$ (selected uniformly at random).
Metropolis-Hastings procedure accepts this candidate partition if its likelihood $p(\paths|\emissionmodel,\hmon,\groups')$ is larger than the likelihood of the current partition $p(\paths|\emissionmodel,\hmon,\groups)$, or, if not, the procedure accepts it with probability equal to the ratio $p(P|\emissionmodel,\hmon,\groups')/p(\paths|\emissionmodel,\hmon,\groups)$.

To compute $p(P|\emissionmodel,\hmon,\groups')$, it is obvious that we do not have to determine the node counts $n_v$ in $\paths$ anew, as they do not depend on the groups.
In contrast, group transition counts depend on the grouping $\groups'$.
However, even the group transitions can be estimated more efficiently than by counting anew from paths $\paths$.
We first do a single pass through paths $\paths$, and determine the node transition counts $n_{v|\bar{v}}$.
For each grouping $\groups$, we then simply use Eq. 13 from the main paper, and sum the counts $n_{v|\bar{v}}$ to obtain the group counts $n_{g|\bar{g}}$.
Thus, we do not pass through the pathway data for every candidate partition, and instead we pass only through the unique transitions between nodes, and map them to transitions between groups.
In summary, to compute the marginal likelihoods, we have to pass only once through the pathway data, and for every partition, we just have to pass through the observed unique transitions of nodes.

\section{All Synthetic Results}
\label{hog:appendix:synthRes}
In \cref{hog:table:synthetic:supplement} we present the full table of synthetic results. 
This table complements \cref{hog:table:synthetic_results} (main text) presenting results for additional node2vec hyperparameters.   

\clearpage

\begin{table}[!ht]
    \centering
    \caption{All Synthetic Results}
    \label{hog:table:synthetic:supplement}
\begin{tabular}{ll|lll}
dataset/runs & Method & $k_\mathcal{H}$ &  AMI & $\sigma$ \\
\hline\hline
    \emph{synth-1} & B-HOG & 1 &    1.0&0.0 \\
      {$100$ runs} & InfoMap & - &  0.11&0.31 \\
      & $\text{node2vec}_{p=1, q=1}$ & - &  0.06&0.27 \\
      & $\text{node2vec}_{p=1, q=4}$ & - &  0.06&0.26 \\
      & $\text{node2vec}_{p=4, q=1}$ & - &  0.04&0.23 \\
      & $\text{node2vec}_{p=4, q=4}$ & - &  0.03&0.21 \\
    \hline
    \emph{synth-2} & B-HOG & 2 &    1.0&0.0 \\
      {$100$ runs} & InfoMap & - &  0.04&0.22 \\
      & $\text{node2vec}_{p=1, q=1}$ & - &  0.02&0.19 \\
      & $\text{node2vec}_{p=1, q=4}$ & - &  0.01&0.17 \\
      & $\text{node2vec}_{p=4, q=1}$ & - &  0.01&0.19 \\
      & $\text{node2vec}_{p=4, q=4}$ & - &  0.01&0.18 \\
    \hline
    \emph{synth-3} & B-HOG & 3 &    1.0&0.0 \\
      {$100$ runs} & InfoMap & - &  -0.0&0.04 \\
      & $\text{node2vec}_{p=1, q=1}$ & - &  0.15&0.28 \\
      & $\text{node2vec}_{p=1, q=4}$ & - &  0.11&0.25 \\
      & $\text{node2vec}_{p=4, q=1}$ & - &   0.04&0.2 \\
      & $\text{node2vec}_{p=4, q=4}$ & - &  0.05&0.18 \\
    \end{tabular}
\end{table}

\vspace{2cm}
\section{All Empirical Results}
\label{hog:appendix:empRes}

In \cref{hog:table:empirical:supplement} we present the full table of empirical results. 
This table complements \cref{hog:table:empirical_results} (in the main text) presenting results for additional node2vec hyperparameters.
\clearpage
\begin{table}[!ht]
    \centering
    \caption{All Empirical Results}
    \label{hog:table:empirical:supplement}

\begin{tabular}{ll|lll}
    dataset & Method & $k_{\hog}$ &  AMI & $\sigma$ \\
    \hline\hline
    \emph{school-11} & B-HOG & 4 & 0.810361 & - \\
     & InfoMap & - & 0.786205 & - \\
     & $\text{node2vec}_{p=1, q=1}$ & - & 0.790080 & 0.014802 \\
     & $\text{node2vec}_{p=1, q=4}$ & - & 0.798534 & 0.002189 \\
     & $\text{node2vec}_{p=4, q=1}$ & - & 0.788449 & 0.014457 \\
     & $\text{node2vec}_{p=4, q=4}$ & - & 0.798439 & 0.003870 \\
    \hline
    \emph{school-12} & B-HOG & 3 & 0.830456	 & - \\
     & InfoMap & - & 0.764800 & - \\
     & $\text{node2vec}_{p=1, q=1}$ & - & 0.955480 & 0.002048 \\
     & $\text{node2vec}_{p=1, q=4}$ & - & 0.955427 & 0.002240 \\
     & $\text{node2vec}_{p=4, q=1}$ & - & 0.955373 & 0.002415 \\
     & $\text{node2vec}_{p=4, q=4}$ & - & 0.955214 & 0.002872 \\
    \hline
    \emph{hospital} & B-HOG & 4 & 0.246311 & - \\
     & InfoMap & - & 0.303180 & - \\
     & $\text{node2vec}_{p=1, q=1}$ & - & 0.098765 & 0.058644 \\
     & $\text{node2vec}_{p=1, q=4}$ & - & 0.075599 & 0.049383 \\
     & $\text{node2vec}_{p=4, q=1}$ & - & 0.105686 & 0.058267 \\
     & $\text{node2vec}_{p=4, q=4}$ & - & 0.091377 & 0.055496 \\
    \hline
    \emph{primary} & B-HOG & 3 & 0.921175 & - \\
     & InfoMap & - & 0.919904 & - \\
     & $\text{node2vec}_{p=1, q=1}$ & - & 0.857491 & 0.015537 \\
     & $\text{node2vec}_{p=1, q=4}$ & - & 0.861175 & 0.005600 \\
     & $\text{node2vec}_{p=4, q=1}$ & - & 0.853804 & 0.017595 \\
     & $\text{node2vec}_{p=4, q=4}$ & - & 0.859867 & 0.009815 \\
    \hline
    \emph{work-13} & B-HOG & 2 & 0.0.758845 & - \\
     & InfoMap & - & 0.758651 & - \\
     & $\text{node2vec}_{p=1, q=1}$ & - & 0.755072 & 0.019565 \\
     & $\text{node2vec}_{p=1, q=4}$ & - & 0.756847 & 0.021343 \\
     & $\text{node2vec}_{p=4, q=1}$ & - & 0.753526 & 0.021522 \\
     & $\text{node2vec}_{p=4, q=4}$ & - & 0.756030 & 0.022311 \\
    \hline
    \emph{work-15} & B-HOG & 2 & 0.671904 & - \\
     & InfoMap & - & 0.716915 & - \\
     & $\text{node2vec}_{p=1, q=1}$ & - & 0.696239 & 0.026103 \\
     & $\text{node2vec}_{p=1, q=4}$ & - & 0.702179 & 0.024661 \\
     & $\text{node2vec}_{p=4, q=1}$ & - & 0.697679 & 0.025749 \\
     & $\text{node2vec}_{p=4, q=4}$ & - & 0.704928 & 0.024739 \\
    \end{tabular}

\end{table}

\section{Optimization with fixed order}
\label{hog:appendix:fixedorder}
We demonstrate the real-world relevance of our method by showing that a higher-order B-HOG obtains a log-likelihood higher than that of the first-order B-HOG in all considered datasets.  
To do this, we optimized the partitions of BHOG models ($10$ runs with $5 \times 10^4$ Metropolis-Hastings iterations each) that are constrained to either the first- or the second-order.
As shown in \cref{hog:table:empirical:supplement_fixedorder}, we find that the second order has higher marginal likelihood---and thus lower description length---than the first-order in all datasets, further highlighting that a higher-order is necessary for modeling the mesoscale dynamics of real-world data.

\begin{table}
\centering
\caption{Comparison between Log-Likelihoods obtained with fixed first and second order for BHOG}
\label{hog:table:empirical:supplement_fixedorder}
\begin{tabular}{l|rr}
     dataset &  order &  Log-Likelihood \\
\hline\hline
     \emph{school-11} &      1 &  -109425.418714 \\
     \emph{school-11} &      2 &  -105295.669616 \\
\hline
     \emph{school-12} &      1 &  -188903.713149 \\
     \emph{school-12} &      2 &  -180142.718849 \\
\hline
      \emph{hospital} &      1 &  -149031.071979 \\
      \emph{hospital} &      2 &  -139812.183362 \\
\hline
    \emph{primary} &      1 &  -845556.115375 \\
    \emph{primary} &      2 &  -788773.924551 \\
\hline       
       \emph{work-13} &      1 &   -32848.199136 \\
       \emph{work-13} &      2 &   -32047.929207 \\
\hline
        \emph{work-15} &      1 &  -327207.508692 \\
        \emph{work-15} &      2 &  -304350.111009 \\
\end{tabular}
\end{table}

\section{Optimal order of ground truth labels}
\label{hog:appendix:optimalordergt}

\begin{figure*}[!ht]
    \centering
    \includegraphics[width = \textwidth]{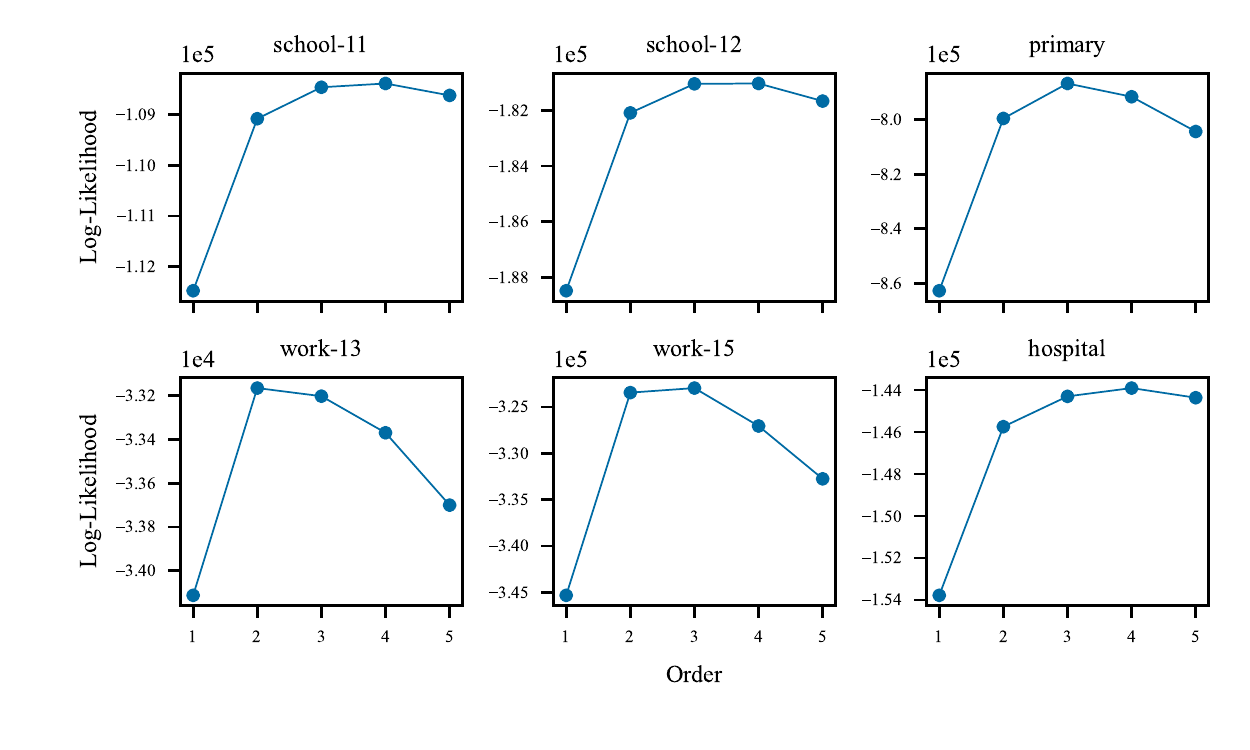}
    \caption{
    Log-likelihood at different orders for the group assignments given by the node's metadata. 
    The optimal order is in the range $2$-$4$ for all datasets, thus highlighting that the optimal description of the mesoscale dynamics from datasets' labels requires accounting for non-Markovian correlations.  
    }
    \label{fig:order_metadata_labels}
\end{figure*}

In this section, we demonstrate the relevance of the higher-order patterns detected by our methods by showing that they are necessary for modeling the dynamics of the mesoscales encoded in the ground truth labels. 
We enforce the ground truth labels as optimal and compute the model's log-likelihood that describes these dynamics at each order.
We observed that the optimal order that best compresses the dynamics is 2-4 in all datasets. 
Specifically $4$ in \emph{school-11}, \emph{school-12}, and \emph{hospital};
$3$ in \emph{primary} and \emph{work-15}; 
$2$ in \emph{work-13} (see \cref{fig:order_metadata_labels}).
Since Bayesian order selection has built-in Occam's razor, these are not spurious correlations. 
Therefore, one cannot use the first-order network model to represent them and, consequently, the mesoscale dynamics encoded in the ground truth labels.

\clearpage
\section{Optimization from ground truth labels}
\label{hog:appendix:optfromgt}
\begin{figure*}[!ht]
    \centering
    \includegraphics[width = \textwidth]{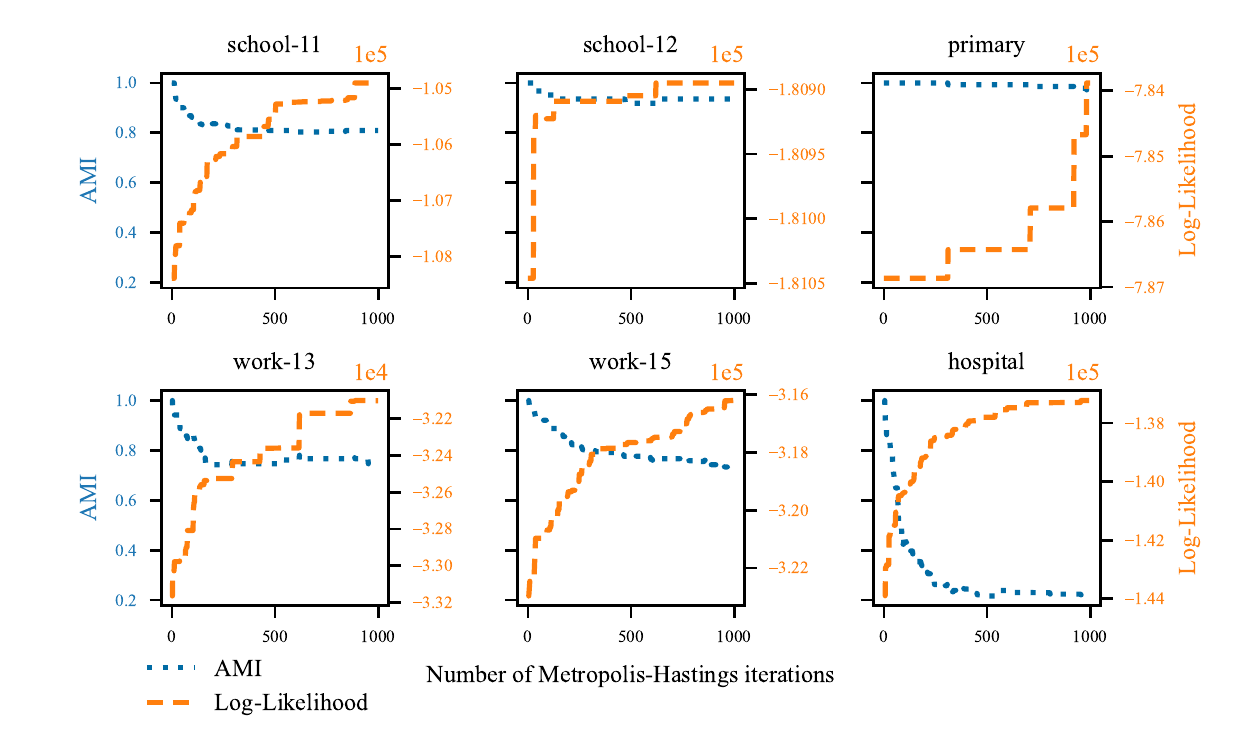}
    \caption{Optimizing Log-Likelihood by starting from the ground truth labels. Notice that the AMI (blue, dotted) goes down while the Log-Likelihood (orange, dashed) goes up.}
    \label{fig:optimize_from_labels}
\end{figure*}

In this section, we show that the mesoscale dynamics identified by BHOG differs from those based on the datasets' labels. 
We use the datasets' labels as a starting point for each dataset and then run $1000$ Metropolis-Hastings iterations. 
As shown in \cref{fig:optimize_from_labels}, with more and more iterations, we always have a decrease in AMI (blue dotted curve) and an increase in Log-Likelihood (orange dashed curve), indicating that the optimization is moving away from initial labels.
The increasing Log-likelihood indicates that, although we are moving away from the datasets labels, we found partitions that compress the data better than the empirical labels.
Therefore, the dynamics in the system (captured by paths) is based on a partition different from the one based on the measured labels.
In conclusion, although our detected partition does not approximate the labels in the data, it better describes the dynamics of complex systems.

\end{document}